\documentclass[aip,pop,amsmath,amssymb,reprint]{revtex4-2}

\usepackage{graphicx}
\usepackage{amsmath}
\usepackage{bm}
\usepackage{color}

\begin{document}

\title{Heat transport in a flowing complex plasma in microgravity conditions}

\author{V. Nosenko}
\email{V.Nosenko@dlr.de}
\affiliation{Institut f\"{u}r Materialphysik im Weltraum, Deutsches Zentrum f\"{u}r Luft- und Raumfahrt (DLR), D-82234 We{\ss}ling, Germany}
\affiliation{Center for Astrophysics, Space Physics, and Engineering Research, Baylor University, Waco, Texas 76798-7310, USA}

\author{S. Zhdanov}
\affiliation{Institut f\"{u}r Materialphysik im Weltraum, Deutsches Zentrum f\"{u}r Luft- und Raumfahrt (DLR), D-82234 We{\ss}ling, Germany}

\author{M. Pustylnik}
\affiliation{Institut f\"{u}r Materialphysik im Weltraum, Deutsches Zentrum f\"{u}r Luft- und Raumfahrt (DLR), D-82234 We{\ss}ling, Germany}

\author{H. M. Thomas}
\affiliation{Institut f\"{u}r Materialphysik im Weltraum, Deutsches Zentrum f\"{u}r Luft- und Raumfahrt (DLR), D-82234 We{\ss}ling, Germany}

\author{A. M. Lipaev}
\affiliation{Institute for High Temperatures, Russian Academy of Sciences, Izhorskaya 13/19, 125412 Moscow, Russia}

\author{O. V. Novitskii}
\affiliation{Gagarin Research and Test Cosmonaut Training Center, 141160 Star City, Moscow Region, Russia}

\date{\today}

\begin{abstract}
Heat transport in a three-dimensional complex (dusty) plasma was experimentally studied in microgravity conditions using Plasmakristall-4 (PK-4) instrument on board the International Space Station (ISS). An extended suspension of microparticles was locally heated by a shear flow created by applying the radiation pressure force of the manipulation-laser beam. Individual particle trajectories in the flow were analysed and from these, using a fluid heat transport equation that takes viscous heating and neutral gas drag into account, the complex plasma's thermal diffusivity and kinematic viscosity were calculated. Their values are compared with previous results reported in ground-based experiments with complex plasmas.
\end{abstract}


\maketitle

\section{Introduction}

Transport phenomena in liquids are important in fundamental science and in engineering applications. They are, however, challenging to study experimentally. To perform experiments at the most fundamental kinetic level (of individual particles), model systems are necessary. A particularly convenient model system is complex plasma. This is a suspension of micron size particles in ionized gas \cite{Ivlev_book}. The particles become charged by collecting electrons and ions from the plasma and interact with each other via a screened Coulomb potential which in many cases can be approximated by the Yukawa potential. The advantages of complex plasmas as model systems are the relative ease of individually observing the constituent particles, stretched spatial and temporal scales, and undamped dynamics. Complex plasmas were successfully used to study diffusion \cite{Nunomura:06Diff}, momentum \cite{Nosenko:2004,Gavrikov:2005,Vorona:2007,Ivlev:2007,Nosenko:2020}, and heat transport \cite{Nunomura:05ThermalCond,Nosenko:08ThermalCond,Fortov:TC}. Heat transport was experimentally studied in a crystalline phase of 2D complex plasma \cite{Nunomura:05ThermalCond}, in a 2D complex plasma at melting conditions \cite{Nosenko:08ThermalCond}, and in a 3D liquid complex plasma \cite{Fortov:TC}. So far, the heat transport in complex plasmas was studied in ground-based laboratory experiments, where the particle suspension was compressed by the force of gravity.

In this paper, we present an experiment with liquid complex plasma, where the heat transport was studied by locally heating the particle suspension by a laser-driven shear flow and analyzing the resultant particle kinetic temperature profile. To achieve an extended unstressed 3D particle suspension, the experiments were performed in microgravity conditions. One advantage of studying transport phenomena in a 3D sample of complex plasma is that the existence of valid transport coefficients is not questioned here, unlike in a 2D sample \cite{Ernst:70}.

\section{Experimental method}

The experiments described in this paper were performed using the PK-4 instrument on board the International Space Station (ISS) \cite{Pustylnik:2017}. The experimental procedure and parameters were similar to those used in Ref.~\cite{Nosenko:2020}. We used DC discharge to produce plasma in Ne at the pressure of $15$~Pa. The discharge current was $0.5$~mA and the maximum voltage was $2.5$~kV. Estimates are available of the electron density $n_e=0.92\times10^8$\,cm$^{-3}$ and temperature $T_e=9.8$~eV (on the tube axis in the middle of the working area) in a similar discharge in Ne at the pressure of $20$~Pa, see Ref.~\cite{Pustylnik:2017} for more details. The screening length in these conditions was $\lambda_D=122~\mu$m. A cloud of melamine formaldehyde (MF) microspheres with a diameter of $3.38\pm0.07$ was suspended in the working area of the discharge tube. They were trapped using polarity switching with a frequency of $500$~Hz and duty cycle $\simeq50$\%. The Wigner-Seitz radius of the particle suspension was calculated as $r_{\rm WS}=r_0/1.79=100~\mu$m, where $r_0$ is the first peak position of the pair correlation function $g(r)$ measured in 2D cross sections of the particle suspension \cite{Liu:2015}. The particle number density was calculated as $n=3(4\pi r_{\rm WS}^3)^{-1}=2.4\times10^5$\,cm$^{-3}$. For the particle charge, we adopted the value $Q=2800e$ reported for our experimental conditions in Ref.~\cite{Antonova:2019}. The screening parameter was $\kappa=r_{\rm WS}/\lambda_D=0.8$.

A shear flow in the particle suspension was created by applying the radiation pressure force from the focused beam of a $808$-nm manipulation laser. The laser beam was aligned with the discharge tube axis and its output power was set to $2.16$~W. As in Ref.~\cite{Nosenko:2020}, the main experimental run was supplemented by a plasma-off phase, where the plasma was briefly (during $0.5$~s) switched off while the manipulation laser remained on. During the plasma-off time, the particle charge rapidly declined \cite{Ivlev:2003}, as did the interparticle interactions and external confinement; however, the particle suspension did not collapse due to weightlessness. The particles were illuminated by a thin sheet of a $532$-nm laser and imaged by two video cameras with slightly overlapping fields of view. Here, we used the video recorded by camera 2. To better resolve the particle motion, a higher than in Ref.~\cite{Nosenko:2020} camera frame rate of $100$ frames per second was used at the expense of narrower field of view (FoV), which measured $1600\times267~{\rm pixels}^2$ or $22.72\times3.79~{\rm mm}^2$. The reduced FoV, however, included the full width of the shear flow with boundary layers. The manipulation-laser beam was coming in the positive $x$ direction, the particles were imaged in the $(x,z)$ plane, and in the scanning direction the imaging system was fixed at $y=0$.

\begin{figure}[tb]
\centering
\includegraphics[width=1.0\columnwidth]{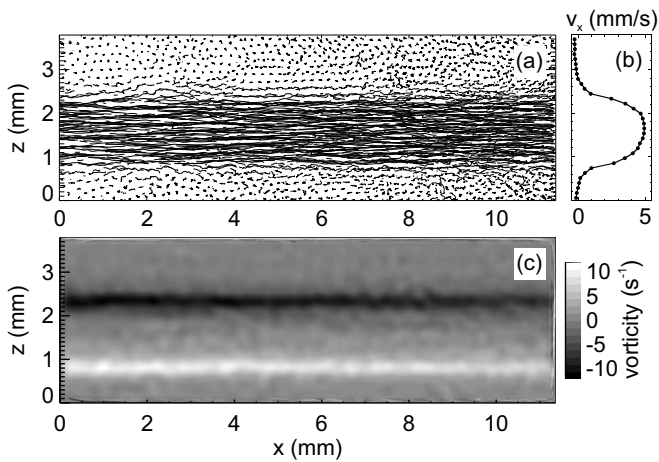}
\caption {\label {Fig_traj_vor} (a) Trajectories of individual particles in the central cross section ($y=0$) of the steady-state laser-induced shear flow during $0.3$~s. The laser radiation force pushes the particles
in the positive $x$ direction. (b) Time-averaged particle velocity profile $v_x(z)$. (c) Map of the time-averaged vorticity $(\nabla \times {\bf v})_y$, which shows the spatial localisation of the shear in the particle flow. In (b) and (c), time averaging was performed during $1.2$~s. Shown here is a part of the original FoV delimited by $400~{\rm pixels}<x<1200~{\rm pixels}$.
}
\end{figure}

Experimental data were analysed using the following method. A $1.2$-s section of the experimental video recording of the steady-state shear flow was identified and used for analysis. Since the shear flow has cylindrical symmetry \cite{Nosenko:2020}, it is sufficient to analyze its central cross section (defined by $y=0$). Particles were identified and traced from frame to frame using a moment method \cite{SPIT}. A part of the original FoV where the particle suspension was rather uniform (delimited by $400~{\rm pixels}<x<800~{\rm pixels}$), was divided into $33$ bins elongated along the $x$ axis and $8$-pixels wide in the $z$ direction. The average number of particles was $\approx 50$ per bin per frame, to the total of $\approx 6000$ per bin for the whole sequence. In each bin, the distributions of the particle velocity $v_{x,z}$ were calculated during $1.2$~s. In all bins, the distributions were close to Maxwellian (shifted for the $v_{x}$ distribution). Therefore, the time-averaged profiles of the particle speed $v_{x,z}(z)$ and kinetic temperature $T_{x,z}(z)$ were calculated. The particle kinetic temperature was calculated as $T_{x,z}=m\langle(v_{x,z}-\overline{v}_{x,z})^2\rangle/k_B$, where $m$ is the particle mass and $k_B$ is the Boltzmann constant. The errors in calculating $v_{x,z}(z)$ and $T_{x,z}(z)$ are estimated as $\pm0.04$~mm/s  and up to $30\%$, respectively.

\section{Experimental results}

The trajectories of individual particles in the central cross section of steady-state shear flow are shown in Fig.~\ref{Fig_traj_vor}(a). The flow appears to be laminar on average with particles performing random motion on top of it. The time-averaged particle velocity profile $v_x(z)$ is shown in Fig.~\ref{Fig_traj_vor}(b). The map of the time-averaged vorticity $(\nabla \times {\bf v})_y$, which shows the spatial localisation of shear in the particle flow, is shown in Fig.~\ref{Fig_traj_vor}(c). The shear is localized in two rather narrow boundary layers adjacent to the flow.

\begin{figure}[tb]
\includegraphics[width=0.9\columnwidth]{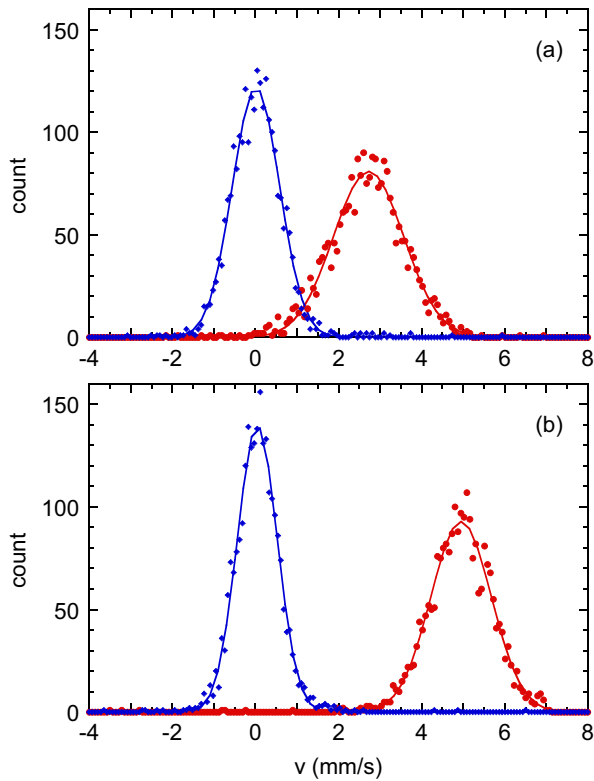}
\caption{\label{Fig_V_distr} Distributions of the particle velocity $v_{x}$ (red circles) and $v_{z}$ (blue diamonds) measured in (a) the area of maximum shear, $0.8~{\rm mm}<z<0.91~{\rm mm}$ and (b) in the flow bulk, $1.7~{\rm mm}<z<1.81~{\rm mm}$. The lines are (shifted) Maxwellian fits.}
\end{figure}

In each narrow bin corresponding to one point in Fig.~\ref{Fig_traj_vor}(b), the distributions of the particle velocity $v_{x,z}$ were close to Maxwellian (shifted for $v_{x}$). Two examples are shown in Fig.~\ref{Fig_V_distr}. The distributions shown in panel (a) were measured in the area of maximum shear, $0.8~{\rm mm}<z<0.91~{\rm mm}$ and those in panel (b) in the flow bulk, $1.7~{\rm mm}<z<1.81~{\rm mm}$. The nearly Maxwellian velocity distributions validate our calculating the mean velocities $\overline{v}_{x,z}$ and also allow us to introduce the notion of the particle kinetic temperature $T_{x,z}=m\langle(v_{x,z}-\overline{v}_{x,z})^2\rangle/k_B$, which describes the random component of the particle motion.

\begin{figure}[tb]
\includegraphics[width=0.9\columnwidth]{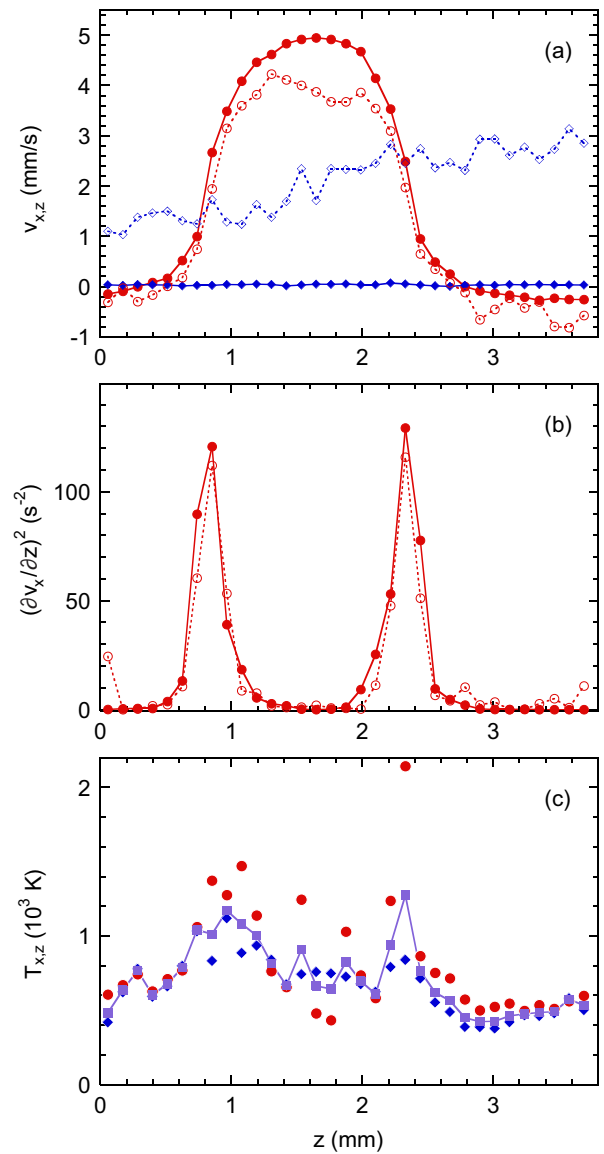}
\caption{\label{Fig_VT} (a) Flow velocity profiles $v_{x,z}(z)$ and (b) squared first derivative $(\partial v_x(z)/\partial z)^2$, to which the heat source is proportional. The red circles are for $v_x(z)$, blue diamonds for $v_z(z)$, solid symbols correspond to the plasma-on phase, open symbols to the plasma-off phase. (c) Longitudinal $T_x(z)$ (red circles), transverse $T_z(z)$ (blue diamonds), and ``thermalized'' $T=(T_x+2T_z)/3$ (pink squares) particle kinetic temperature profiles for the plasma-on phase.}
\end{figure}

The time-averaged flow velocity profiles $v_{x,z}(z)$ are shown in Fig.~\ref{Fig_VT}(a). In the plasma-on phase (shown by the solid lines), the $v_{x}(z)$ component is approximately symmetric with respect to $z=1.65$~mm, while the $v_{z}(z)$ component is negligibly small, which supports the assumption of cylindrical symmetry of the flow. In the plasma-off phase (shown by the dashed lines), the $v_{x}(z)$ component does not change much (except in the central part). However, the $v_{z}(z)$ component increases substantially due to the drift of the particle cloud in the positive $z$ direction (presumably due to the thermophoretic force). These observations corroborate the findings of Ref.~\cite{Nosenko:2020}. Since the source function of viscous heating in the flow geometry of our experiment is proportional to $\eta(\partial v_x(z)/\partial z)^2$, where $\eta$ is the shear viscosity of the particle suspension \cite{Landau}, it is instructive to plot $(\partial v_x(z)/\partial z)^2$ as shown in Fig.~\ref{Fig_VT}(b). This figure serves to visualize the distribution of heat sources in the shear flow.

The time-averaged particle kinetic temperature profiles in the plasma-on phase $T_{x,z}(z)$ are shown in Fig.~\ref{Fig_VT}(c). The temperature is anisotropic with its longitudinal component greater on average than transverse, $T_x(z)>T_z(z)$. In what follows, we will use the ``thermalized'' $T=(T_x+2T_z)/3$. There are two prominent peaks in $T(z)$ corresponding to the viscous heating sources [cf. Fig.~\ref{Fig_VT}(b)]. This is a direct observation of viscous heating in a flowing liquid 3D complex plasma, which was previously reported only in 2D complex plasmas \cite{Nosenko:2004,Feng:2012}. In the plasma-off phase, the random part of the particle motion was very strong and it was not possible to calculate their kinetic temperature.

In very simplified terms, the main idea of the subsequent analysis is to deduce the shear viscosity $\eta$ of the particle suspension from the height of the temperature peaks (since the heat source function is proportional to $\eta$) and its thermal conductivity from the peak widths.

\section{Heat transport model}

The cylindrical symmetry of the flow significantly simplifies theoretical consideration. In what follows, we will base our model on this suggestion assuming that the main flow parameters are only dependent on the cylindrical radius $r$, that is, $T=T(r), v_x= v_x(r)$, etc. Moreover, when plasma is on and the radial confinement suppresses well the radial particle motion, the radial flow component is small compared to the axial one, see Fig. \ref{Fig_VT}(a), therefore, it can be ignored in the model.

Generally, when considering heat transport in any system, one has to address two major points: (i) the source of heat, (ii) the mechanism of heat redistribution. In complex plasmas, the heat transport is traditionally characterised studying the particle kinetic temperature distributions \cite{Nunomura:05ThermalCond,Nosenko:08ThermalCond,Fortov:TC}. It is convenient to classify the sources of heat as local ones and those globally distributed. We attribute the quite intense local heating observed in the present experiment to the internal \emph{viscous heating} \cite{Landau}. It is well evidenced by the distribution of the squared derivative of the particle flow velocity, to which this type of heat source is proportional in our case. It is sharply peaked in the vicinity of the inflection point of the flow velocity profile, see Fig.~\ref{Fig_VT}(b). Noticeably, the kinetic temperature also demonstrates prominent maxima in the vicinity of these points, see Fig.~\ref{Fig_VT}(c). In contrast, the distributed heat sources are spread approximately uniformly inside the particle cloud. They can be attributed to external sources, e.g., those stemming from the thermophoretic force caused by the non-uniformly heated discharge tube walls \cite{Rothermel:2002} or the photophoretic force from the manipulation laser \cite{Nosenko:2020JP}.

In what follows, we demonstrate that the assumptions based on these experimental observations significantly simplify the analysis of the heat transport and allow us to come, with relatively simple means, to rather important conclusions.

\subsection{The governing equations}

The heat is transported within the particle cloud by mutual collisions which are interpreted here as the thermal conductivity. Under steady-state conditions of the particle flow with cylindrical symmetry, this process can be described by the following simple model:

\begin{equation}\label{1}
b^2\Delta T \equiv b^2\frac{d}{rdr}(r\frac{dT}{dr})= T-T_{\rm bg}-S,
\end{equation}
where $T$ is the ``thermalized'' particle kinetic temperature, $T_{\rm bg}={\rm const}$ is the background kinetic temperature due to the distributed heat sources, while $b$ is the heat transport length defined by the cloud thermal diffusivity $\chi$ via the relationship
\begin{equation}\label{2}
\chi=\frac{2\gamma b^2}{c_p},
\end{equation}
where $\gamma$ is the Epstein neutral gas drag rate \cite{Epstein:1924} and $c_p=5/2$ is the specific heat capacitance. Finally, $S=S(r)$ is the local heat source  proportional to the squared derivative of the flow velocity $v'^2$, the particle mass $m$, and the cloud kinematic viscosity $\nu=\eta(mn)^{-1}$:
\begin{equation}\label{3}
S=Cv'^2(r)\equiv \frac{m \nu }{4\gamma}v'^2(r).
\end{equation}
The values $b$, $T_{\rm bg}$, and $C$ are constant parameters that have to be obtained by fitting the solution of the model (\ref{1}),(\ref{2}),(\ref{3}) to experimental data. An advantage of this model is that Eq.~(\ref{1}) can be solved analytically:
\begin{equation}\label{4}
\begin{split}
T=T(r)=T_{\rm bg}+I_0\left(\frac{r}{b}\right)\int_r^{\infty}\frac{d\xi \xi}{b^2}S(\xi)K_0\left(\frac{\xi}{b}\right)+ \\
K_0\left(\frac{r}{b}\right)\int_0^r\frac{d\xi \xi}{b^2}S(\xi)I_0\left(\frac{\xi}{b}\right),
\end{split}
\end{equation}
where $I_0$ and $K_0$ are the modified Bessel functions of order zero. Note an important consequence of this relationship:
\begin{equation}\label{5}
\int_0^{\infty}d\xi \xi \left[T(\xi)-T_{\rm bg}\right]=C\int_0^{\infty}d\xi \xi v'^2(\xi).
\end{equation}
Actually, the integrals in Eq.~(\ref{5}) can be measured experimentally (of course, using some crude approximation for $T_{\rm bg}$), which allows one to estimate $C$ and then viscosity by using Eq.~(\ref{3}). This may serve as a first step when fitting the analytically computed kinetic temperature profiles to those experimentally measured.

\subsection{Further simplifications}

Since the exact analytical solution (\ref{4}) is somewhat cumbersome, it is worth considering further possible simplifications. Given quite narrow and sharply peaked distribution of the heating term $S(r)$ [see Fig.~\ref{Fig_VT}(b)], Eq.~(\ref{4}) can be reduced to the following form, which is well suited for fitting purposes:
\begin{equation}\label{6}
T=T(r)=T_{\rm bg}+A\tau (r), \tau = \left\{
                                \begin{array}{ll}
                                  I_0\left(\frac{r}{b}\right)/I_0\left(\frac{R}{b}\right), & \hbox{r $\leq$ R;} \\
                                  K_0\left(\frac{r}{b}\right)/K_0\left(\frac{R}{b}\right), & \hbox{r $>$ R,}
                                \end{array}
                              \right.
\end{equation}
where the parameter $R$, considered fixed in what follows, corresponds to the radial coordinate of the inflection point well pronounced in the flow velocity distribution, see Figs.~\ref{Fig_VT}(a),(b), while the parameter $A$ stands for
\begin{equation}\label{7}
A=\frac{C}{b^2}\int_0^{\infty}d\xi \xi v'^2(\xi)
\end{equation}
and defines the excess of the maximum kinetic temperature $T_{\rm max}$ over the background value $T_{\rm bg}$. The expression  $T_{\rm max}=T(R)=T_{\rm bg}+A$ becomes exact at $r=R$. Note that $T_{\rm bg}$, $A$, and $\mathcal{R}=R/b$ represent an optimal set of fitting parameters.

\subsection{The scoring function}

As a scoring function for performing fitting iterations, it is advantageous to introduce
\begin{equation}\label{8}
f=f(T_{\rm bg},A, \mathcal{R}) = \left<\left(T(r)-T_{\rm bg}-A \tau(r)\right)^2\right>\Rightarrow \min,
\end{equation}
where $\left<...\right>$ denotes the average over available range of $r$. It is convenient to introduce also the mean values $\overline{T}=\left<T\right>$, $\overline{\tau}=\left<\tau\right>$ and for the sake of brevity the deviations from the mean values $\widetilde{T}=T-\overline{T}$, $\widetilde{\tau}=\tau-\overline{\tau}$; the scoring function then reduces to
\begin{equation}\label{9}
\begin{split}
f= \left(\overline{T}-T_{\rm bg}-A\overline{\tau}\right)^2-\frac{\left<\widetilde{\tau}\widetilde{T}\right>^2}{\left<\widetilde{\tau}^2\right>}+\left<\widetilde{T}^2\right>+ \\ \left<\widetilde{\tau}^2\right>\left(A-\frac{\left<\widetilde{T} \widetilde{\tau}\right>}{\left<\widetilde{\tau}^2\right>}\right)^2,
\end{split}
\end{equation}
with the evident minimum/optimal essentially positive value
\begin{equation}\label{10}
f=f^{\rm opt}(\mathcal{R})=\left<\widetilde{T}^2\right>-\frac{\left<\widetilde{\tau}\widetilde{T}\right>^2}{\left<\widetilde{\tau}^2\right>},
\end{equation}
which is reached at
\begin{equation}\label{11}
A^{\rm opt}=\frac{\left<\widetilde{T} \widetilde{\tau}\right>}{\left<\widetilde{\tau}^2\right>},~T_{\rm bg}^{\rm opt}=\overline{T}-A^{\rm opt}\overline{\tau}.
\end{equation}

\begin{figure}
\includegraphics[width=0.9\columnwidth]{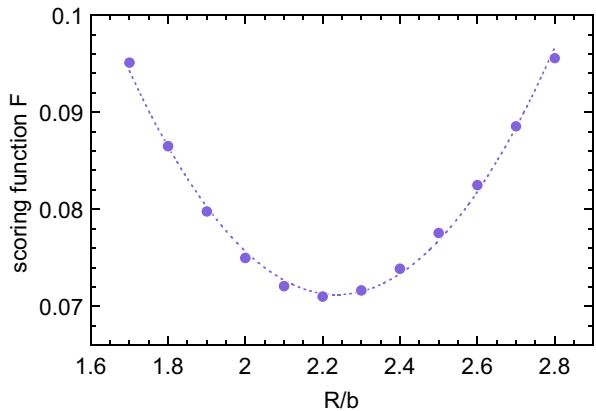}
\caption{\label{Fig_F} Normalized scoring function $F=\langle(T_{\rm exp}-T_{\rm fit})^2\rangle/\langle(T_{\rm exp}-\langle T_{\rm exp}\rangle)^2$ with second-order polynomial fit.}
\end{figure}

\begin{figure}
\includegraphics[width=0.9\columnwidth]{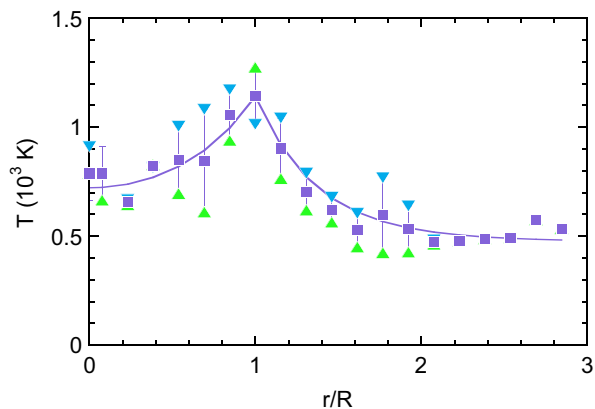}
\caption{\label{Fig_Tfit} Particle kinetic temperature profile $T(r)$ for the plasma-on phase with theoretical fit using Eq.~\ref{6}. Blue down triangles are for $z<1.59$~mm, green up triangles for $z>1.59$~mm, and purple squares are their mean values.
}
\end{figure}

\begin{table}\label{table:1}
\caption{Fitting parameters, see the text for details.}
\begin{tabular}{|c|c|c|c|c|c|}
  \hline
  $R/b$         & $A$~(K)    & $T_{\rm bg}$~(K)  & $T_{\rm max}^{\rm exp}$~(K) & $T_{\rm max}^{\rm fit}$~(K) & $R$~(mm) \\
  \hline
  $2.24\pm0.11$ & $665\pm22$ & $476\pm16$        & $1144\pm131$                & $1141\pm38$                 & 0.738    \\ \hline
\end{tabular}
\end{table}

Finally, it is enough to further minimize the scoring function already optimized over the $A$ and $T_{\rm bg}$ parameters, see Eq.~(\ref{10}), which is quite a straightforward task, see Fig. \ref{Fig_F}. This procedure results in the optimized set of parameters shown in Table~I. Together with Eq.~(\ref{6}), they are sufficient to successfully solve the fitting problem for our experimental data, see Fig.~\ref{Fig_Tfit}.

\section{Thermal diffusivity and kinematic viscosity}

An estimate of the heat transport length follows immediately from $b=R/\mathcal{R}$ and then, with $b$ known, the thermal diffusivity can be calculated using Eq.~(\ref{2}). Both estimates are collected in Table~II. A fairly high accuracy, despite quite large scatter of the experimental points, of these estimates is worth mentioning. To estimate the kinematic viscosity, note first that from the definitions Eqs.~(\ref{4}),(\ref{6}) it follows
\begin{equation}\label{12}
A=Cb^{-2}I_0(\mathcal{R})K_0(\mathcal{R})\int_0^{\infty}d\xi \xi v'^2(\xi).
\end{equation}
Further, given that $C=\frac{m \nu }{4\gamma}$, the kinematic viscosity is easy to estimate. The result is shown in Table~II. The momentum transport scale length $a$ is also shown there. Its value follows immediately from the simple scaling relationship $\nu = a^2 \gamma$. Note that, compared to the heat transport parameters, the latter estimates are significantly less accurate. The source of the inaccuracy is the quite shallow ``scoring well'', see Fig. \ref{Fig_F}.

\begin{table}\label{table:2}
\caption{Estimated complex-plasma's thermal diffusivity $\chi$, kinematic viscosity $\nu$, momentum transport length $a$, heat transport length $b$, mean free path $\ell$, and energy exchange rate $\beta$.}
\begin{tabular}{|c|c|c|c|c|c|}
  \hline
  $\chi ({\rm mm}^2$/s) & $\nu ({\rm mm}^2$/s) & $a$ (mm)      & $b$ (mm)      & $\ell$ (mm) & $\beta ({\rm s}^{-1})$ \\
  \hline
  $3.3\pm0.1$           & $0.8\pm0.4$          & $0.15\pm0.03$ & $0.33\pm0.02$ & $0.01$      & $203$                  \\
  \hline
\end{tabular}
\end{table}

The obtained value of thermal diffusivity $\chi=3.3\pm0.1~{\rm mm}^2$/s is between the values
of $\approx 1~{\rm mm}^2$/s and $\approx 9$~mm$^2$/s reported for a 3D liquid complex plasma \cite{Fortov:TC} and a 2D complex plasma near melting transition \cite{Nosenko:08ThermalCond}, respectively, both in ground-based experiments. The calculated value of kinematic viscosity $\nu=0.8\pm0.4~{\rm mm}^2$/s is consistent with the estimate reported in microgravity experiments \cite{Nosenko:2020}. On the other hand, it is much lower than was previously reported in ground-based experiments with 3D complex plasmas \cite{Gavrikov:2005,Vorona:2007,Ivlev:2007}. In Ref.~\cite{Nosenko:2020}, this difference was attributed to the formation of strings in the particle cloud. The strings tend to align themselves with the axis of discharge tube \cite{Pustylnik:2020,Schwabe:2020,Mitic:2021} and can reduce the viscosity of complex plasma. It is interesting to note that our result for kinematic viscosity of a 3D complex plasma is comparable to that of a 2D complex plasma \cite{Nosenko:2004} $\nu_{\rm 2D}\simeq2~{\rm mm^2/s}$ and also of liquid water \cite{Morfill:04}, $\nu_w\simeq1.8~{\rm mm^2/s}$.

\section{Energy exchange between the kinetic temperature components}

An advantage of working with the ``thermalized'' particle kinetic temperature is that the energy exchange terms are excluded from the heat transport equation. Since the rate coefficient of the energy exchange is {\it apriory} unknown, this approach significantly simplifies the consideration. Still, it would be incorrect to ignore the exchange effects because the observed kinetic temperature components are noticeably different, see Fig.~\ref{Fig_VT}(c).

The ``thermalized'' kinetic temperature is by definition $T=\frac13 \left(T_x+T_y+T_z \right)$, where $T_{x,y,z}=m\left<\left(v_{x,y,z}-\left<v_{x,y,z}\right>\right)^2\right>$ are actually the mean squared random velocity components. In our cylindrically symmetric case, $T_y=T_z$ and this relationship simplifies to $T=\frac13\left(T_{\|}+2T_{\bot}\right)$, where $T_{\|}$ and $T_{\bot}=\frac12\left(T_y+T_z\right)$ are the longitudinal and transverse kinetic temperature components with respect to the particle flow direction $x$.

Let us consider the far-field region away from the intense viscous heating, $r>1.4$~mm. In this region, the kinetic temperature components approach nearly constant mean values, $\left<T\right>=730\pm70$~K, $\left<T_{\|}\right>=1100\pm100$~K, $\left<T_{\bot}\right>=550\pm70$~K. The relaxed mean value, e.g., for $\left<T_{\bot}\right>$ should obey a balance equation
\begin{equation}\label{13}
nc_p\frac{\partial}{\partial t}\left<T_{\bot}\right>=0=-2n\gamma \left<T_{\bot}\right> -n\beta \left(\left<T_{\bot}\right>-\left<T\right>\right) +S_{\bot},
\end{equation}
where $\beta$ is the energy exchange rate defined as the inverse of the energy exchange time, $\tau_{\varepsilon}=\beta^{-1}$, and $S_{\bot}$ symbolises the (minor) external heating. Ignoring this factor, we get
\begin{equation}\label{14}
\beta=6\gamma \frac{\left<T_{\bot}\right>}{\left<T_{\|}\right>- \left<T_{\bot}\right>} \cong 203~\text{s}^{-1}.
\end{equation}

\section{Concluding remarks}

In conclusion, it is worth to discuss the validity of the simplifications made in the heat transport model. First, the applicability of the fluid dynamics approach is justified in the following way: Since the typical thermal velocity of the particles is about $v_T\approx 1$~mm/s, for $\tau_{\varepsilon}\sim1/200$~s the mean free path is of the order of $\ell=2v_T\tau_{\varepsilon}\sim 10^{-2}$~mm, which is much smaller than all other length scales in the problem. Second, the validity of the analysis using simplified Eq.~(\ref{6}) follows from the fact that the momentum transport length $a$ turns out to be smaller than the heat transport length $b$, see Table~II. It is also clear from Figs.~\ref{Fig_VT}(b),(c), that the temperature distribution is wider than the heat source function (proportional to the squared derivative of the flow velocity).

\section{Acknowledgments}

The authors gratefully acknowledge the joint ESA-Roscosmos experiment ``Plasmakristall-4'' on board the International Space Station. This work was supported in part by DLR/BMWi Grant No. 50WM1441. We thank Ch. Knapek for carefully reading the manuscript and helpful comments.

\section{Author declarations}

The authors have no conflicts of interest to disclose.

\section{Data availability}

Raw data were generated at the PK-4 facility on the International Space Station. Derived data supporting the findings of this study are available from the corresponding author upon reasonable request.



\begin{thebibliography}{}

\bibitem{Ivlev_book} A. Ivlev, H. L\"{o}wen, G. Morfill, C. P. Royall, {\it Complex Plasmas and Colloidal Dispersions: Particle-resolved Studies of Classical Liquids and Solids}, Series in Soft Condensed Matter Vol. 5 (World Scientific, Singapore, 2012).

\bibitem{Nunomura:06Diff} S.~Nunomura, D.~Samsonov, S.~Zhdanov, and G.~Morfill, Phys. Rev. Lett. {\bf 96}, 015003 (2006).

\bibitem{Nosenko:2004} V. Nosenko and J. Goree, Phys. Rev. Lett. {\bf 93}, 155004 (2004).

\bibitem{Gavrikov:2005} A. Gavrikov, I. Shakhova, A. Ivanov, O. Petrov, N. Vorona, and V. Fortov, Phys. Lett. A {\bf 336}, 378 (2005).

\bibitem{Vorona:2007} N. A. Vorona, A. V. Gavrikov, A. S. Ivanov, O. F. Petrov, V. E. Fortov, and I. A. Shakhova, Journal of Experimental and Theoretical Physics {\bf 105}, 824 (2007).

\bibitem{Ivlev:2007} A. V. Ivlev, V. Steinberg, R. Kompaneets, H. H\"{o}fner, I. Sidorenko, and G. E. Morfill, Phys. Rev. Lett. {\bf 98}, 145003 (2007).

\bibitem{Nosenko:2020} V. Nosenko, M. Pustylnik, M. Rubin-Zuzic, A. M. Lipaev, A. V. Zobnin, A. D. Usachev, H. M. Thomas, M. H. Thoma, V. E. Fortov, O. Kononenko, and A. Ovchinin, Phys. Rev. Research {\bf 2}, 033404 (2020).

\bibitem{Nunomura:05ThermalCond} S.~Nunomura, D.~Samsonov, S.~Zhdanov, and G.~Morfill, Phys. Rev. Lett. {\bf 95}, 025003 (2005).

\bibitem{Nosenko:08ThermalCond} V. Nosenko, S. Zhdanov, A. V. Ivlev, G. Morfill, J. Goree, and A. Piel, Phys. Rev. Lett. {\bf 100}, 025003 (2008).


\bibitem{Fortov:TC} V. E. Fortov, O. S. Vaulina, O. F. Petrov, M. N. Vasiliev, A. V. Gavrikov, I. A. Shakova, N. A. Vorona, Yu. V. Khrustalyov, A. A. Manohin, and A. V. Chernyshev, Phys. Rev. E {\bf 75}, 026403 (2007).

\bibitem{Ernst:70} M. H.~Ernst, E. H.~Hauge, and J. M. J. van Leeuwen, Phys. Rev. Lett. {\bf 25}, 1254 (1970).

\bibitem{Pustylnik:2017} M. Y. Pustylnik, M. A. Fink, V. Nosenko, T. Antonova, T. Hagl, H. M. Thomas, A. V. Zobnin, A. M. Lipaev, A. D. Usachev, V. I. Molotkov, O. F. Petrov, V. E. Fortov, C. Rau, C. Deysenroth, S. Albrecht, M. Kretschmer, M. H. Thoma, G. E. Morfill, R. Seurig, A. Stettner, V. A. Alyamovskaya, A. Orr, E. Kufner, E. G. Lavrenko, G. I. Padalka, E. O. Serova, A. M. Samokutyayev, and S. Christoforetti, Rev. Sci. Instrum. {\bf 87}, 093505 (2016).

\bibitem{Liu:2015} B. Liu, J. Goree, and W. D. Suranga Ruhunusiri, Rev. Sci. Instrum. {\bf 86}, 033703 (2015).

\bibitem{Antonova:2019} T. Antonova, S. A. Khrapak, M. Y. Pustylnik, M. Rubin-Zuzic, H. M. Thomas, A. M. Lipaev, A. D. Usachev, V. I. Molotkov, and M. H. Thoma, Phys. Plasmas {\bf 26}, 113703 (2019).

\bibitem{Ivlev:2003} A. V. Ivlev, M. Kretschmer, M. Zuzic, G. E. Morfill, H. Rothermel, H. M. Thomas, V. E. Fortov, V. I. Molotkov, A. P. Nefedov, A. M. Lipaev, O. F. Petrov, Yu. M. Baturin, A. I. Ivanov, and J. Goree, Phys. Rev. Lett. {\bf 90}, 055003 (2003).

\bibitem{SPIT} U. Konopka ``Super Particle Identification and Tracking'' (unpublished).

\bibitem{Landau} L. D.~Landau and E. M.~Lifshitz, {\it Fluid Mechanics} (Butterworth-Heinemann, Boston, 1997), Vol. 6.

\bibitem{Feng:2012} Y. Feng, J. Goree, and B. Liu, Phys. Rev. E {\bf 86}, 056403 (2012).

\bibitem{Rothermel:2002} H. Rothermel, T. Hagl, G. E. Morfill, M. H. Thoma, and H. M. Thomas, Phys. Rev. Lett. {\bf 89}, 175001 (2002).

\bibitem{Nosenko:2020JP} V. Nosenko, F. Luoni, A. Kaouk, M. Rubin-Zuzic, H. Thomas, Phys. Rev. Research {\bf 2}, 033226 (2020).

\bibitem{Epstein:1924} P. Epstein, Phys. Rev. {\bf 23}, 710 (1924).

\bibitem{Pustylnik:2020} M. Y. Pustylnik, B. Klumov, M. Rubin-Zuzic, A. M. Lipaev, V. Nosenko, A. D. Usachev, A. D. Zobnin, V. I. Molotkov, G. Joyce, H. M. Thomas, M. H. Thoma, O. F. Petrov, V. E. Fortov, and O. Kononenko, Phys. Rev. Res. {\bf 2}, 033314 (2020).

\bibitem{Schwabe:2020} M. Schwabe, S. A. Khrapak, S. K. Zhdanov, M. Y. Pustylnik, C. R\"{a}th, M. Fink, M. Kretschmer, A. M. Lipaev, V. I. Molotkov, A. S. Schmitz, M. H. Thoma, A. D. Usachev, A. V. Zobnin, G. I. Padalka, V. E. Fortov, O. F. Petrov, and H. M. Thomas, New J. Phys. {\bf 22}, 083079 (2020).

\bibitem{Mitic:2021} S. Mitic, M. Y. Pustylnik, D. Erdle, A. M. Lipaev, A. D. Usachev, A. V. Zobnin, M. H. Thoma, H. M. Thomas, O. F. Petrov, V. E. Fortov, and O. Kononenko, Phys. Rev. E {\bf 103}, 063212 (2021).

\bibitem{Morfill:04} G. E. Morfill, M. Rubin-Zuzic, H. Rothermel, A. V. Ivlev, B. A. Klumov, H. M. Thomas, U. Konopka, V. Steinberg, Phys. Rev. Lett. {\bf 92}, 175004 (2004).

\end{thebibliography}
\end{document}